# Compact freeform illumination design by deblurring the extended sources

*Shili Wei[1], Zhengbo Zhu[1], Wenyi Li[1], and Donglin Ma[1,2,*]*

[1]*School of Optical and Electronic Information and Wuhan National Laboratory for Optoelectronics, Huazhong University of Science and Technology, Wuhan, Hubei 430074, China*
[2]*Shenzhen Huazhong University of Science and Technology, Shenzhen 518057, China*
[*]*Corresponding author:* madonglin@hust.edu.cn

**Abstract:** Illumination is the deliberate utilization of light to realize practical or aesthetic effects. The designers combine with the environmental considerations, energy-saving goals, and technology advances with fundamental physics to develop lighting solutions to satisfy all of our ever-changing needs. Achieving highly efficient and precise control of the energy output of light sources while maintaining compact optical structures is the ultimate goal of illumination design. To realize miniaturized and lightweight luminaires, the design process must consider the extents of light sources. However, the illumination design for extended sources is still a challenging and unsolved problem. Here, we propose a method to design ultra-performance illumination optics enabled by freeform optical surfaces. The proposed method is very general with no limitations of far-field approximation and Lambertian luminescent property. We demonstrate the feasibility and efficiency of the proposed method by designing several freeform lenses realizing accurate and highly efficient illumination control as well as ultra-compact structures.

**Introduction**

Freeform optical surfaces have brought substantial changes in a 130-year-old area of optical design including imaging [1] and non-imaging concerns [2]. The unrestricted geometry of optical surfaces enables the possibility of compact, lightweight, and efficient lighting systems with superior optical performance. The basic problem of illumination design is to solve a single surface or multiple surfaces that can regulate the energy output of light sources into a prescribed irradiance distribution on a target surface, which is also one of the main concerns of non-imaging optics [3]. Extensive research has been done into designing freeform lenses for ideal sources (zero-étendue sources) [4–17] enabling to convert the light output into almost any desired irradiance distribution when the source size is negligible compared to the optic. However, for the real sources that have finite extents, the design methods for ideal sources require to design lenses with their sizes several times larger than the sources (usually more than five times) [18], which deviates from the goals of compactness and light-weight for illumination design.

For purpose of obtaining miniaturized illumination lenses, the design process must consider the spatial and angular extents of light sources. However, there is a trade-off between the compactness of freeform lens and the accuracy of illumination control. The design difficulty will increase steeply when the sizes of lenses get close to the sizes of sources, and the precise control of light flux propagation becomes intractable. The compactness of freeform lenses is usually defined as a quantity $h/d$, where $h$ denotes the maximum height of lens and $d$ represents the maximum size of source (e.g. the diameter of a disk source or the diagonal of a square source). As a rule of thumb, the smaller the $h/d$ is, the harder the illumination design becomes in most situations.

Two common methods are popular to tackle the "extended sources" problem including the optimization method [19–22] and the feedback method [23–25]. However, the feedback method easily falls into the local minimum and does not perform well in compact designs with low $h/d$. Besides, the illumination design is a highly non-convex optimization problem and contains too many variables, which causes the global minimum difficult to approach. The optimization has already obtained some excellent results presented by Byzov [22] transforming the light from a square source into uniform square patterns with *h/d* as small as 1.6. Nevertheless, the optimization process is a generalized engineering design method that lacks physical insight for illumination design.

Several direct methods have been proposed based on PDEs and edge rays tailoring but only applicable for 2D cases or rotational geometry [26–29]. There are some other solutions for extended sources such as the phase space formulations [30] and iteratively surface correction with a series of approximations [31]. The most significant work in recent years is the so-called wavefront tailoring method (WTM) [32] for solving freeform surfaces delivering a prescribed far-field irradiance distribution. The method transforms the design problem into the wavefronts coupling problem [32, 33] and solve the optics based on the SMS3D method [34]. The WTM provides physical insights in the field of illumination design by revealing the correspondence of wavefront propagation and the spatial energy distribution of extended sources. However, the recent WMT still has limitations of its application to non-Lambertian sources, near field illumination (the size of optic cannot be ignored comparing to the target), and various shaped targets and sources.

As mentioned above, the freeform illumination design for extended sources is still not well addressed and requires further research. In this paper, we propose a method that considers the illumination of extended sources as an integral of the irradiance from an ideal source and the blur from the extent of the light source. Some researches share similar ideas but unfortunately are only applicable to extended sources with negligible sizes [33, 34] or relying on additional optical system [35]. In this paper, we reveal that the blur patterns caused by the extension of light sources can be analytically determined based on the basic calculation of radiometry and ray tracing. This transforms the illumination design for extended sources to an ideal source problem combined with a spatially variant deconvolution process. The proposed method is very general that can tailor the illumination design for both far-field, near-field targets, Lambertian, and non-Lambertian sources with any shapes. We designed several freeform lenses with high efficiency and compactness as well as precise control of light rays.

**Method**

**Basic theory**

We consider the illumination system combining a light source, a freeform lens, and a plane target as shown in Fig. 1. The origin is located at the center of the extended source, where we assume that a virtual point light source is shining here. For a zero-étendue source, each ray of the incident beam can be determined by two parameters $\mathbf{u} = (u, v)$ [36]. The predefined target irradiance $E_p(x, y)$ will directly determine a freeform surface. And we can regard the freeform surface as a parametric surface $\mathbf{f}$ with parameters on target coordinates $(x, y)$. The freeform illumination design for ideal sources is very mature that can transform the light into almost any irradiance distributions $E_p(x, y)$ on target plane. As shown in Fig. 1(a), each point on target plane will correspond to one single ray and produce a pulse irradiance function.

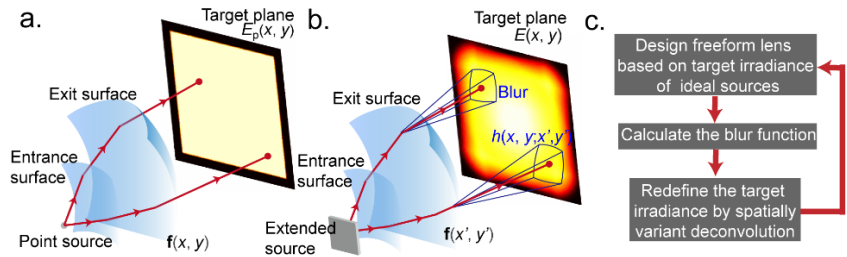

Fig. 1 (a) Illumination design of an ideal source; (b) illumination from extended sources; (c) the connection of illumination design between ideal sources and extended sources; (d) design procedure for extended sources.

The basic idea of this work is considering the illumination from extended sources as a blurring effect on the irradiances from idea sources. As shown in Fig. 1(b), the effect of extended sources can be regarded as angular extents of the rays from the exit surface, which will produce a blurring pattern on the target instead of an impulse from the ideal source. For each point on a target point $(x, y)$ as well as the freeform surface $\mathbf{f}(x, y)$, the extended source will produce a blur function $h(x, y; x', y')$ on target plane. The irradiance $E(x, y)$ can be formulated as an integral transform:

$$E(x, y) = \iint h(x, y; x', y') E_p(x', y') dx' dy'. \tag{1}$$

Eq. (1) demonstrates that the irradiance can be calculated by an integral of the blur kernel and the irradiance from ideal sources. Each point $(x, y)$ in target plane will correspond to a freeform surface point $\mathbf{f}(x, y)$ and a blur kernel function $h(x, y; x', y')$.

Based on the above considerations, we can formulate a design process of an illumination system for extended sources. If the blur kernel function has been already known, we can use the prescribed irradiance to calculate the target irradiance of an ideal source by a spatially variant deconvolution which is an inverse process of Eq. (1). Then, the illumination design for extended sources will be transformed into the design process of ideal sources. However, the blur kernel of the final system cannot be directly estimated. Therefore, an initial freeform lens is designed based on ideal source assumption and the blur kernel should be evaluated. The calculation of the blur kernel function will be discussed in the next section. Then we can redefine the target irradiance of ideal sources based on a spatially variant deconvolution. In this step, the mapping m is assumed to be unchanged. The process will be iterated until obtaining a satisfactory result. The design procedure is shown in Fig. 1(c). Notice that any well-developed zero-étendue algorithm can be employed or modified to adapt our design framework. In this paper, we will employ the least-squares ray mapping (LSRM) method proposed in our previous work [17] to accomplish the illumination design.

**Calculation of the blur kernel**

The calculation of the 4D blur kernel function seems to be an intractable task. In this section, we will demonstrate that the blur kernel function can be analytically calculated.

We first consider the basic analysis of flux transfer shown in Fig. 2(a). The ray with radiance $L$ emerges from an elementary source surface $ds_1$ and travels to an elementary receiving surface $ds_2$. The irradiance of $ds_2$ produced by $ds_1$ is given by [39]:

$$dE = L \frac{\cos\theta_1 \cos\theta_2}{R^2} ds_1, \tag{2}$$

where $\theta_1$, $\theta_2$ denote the angle made by the direction of the ray with respect to the normals of the source surface and the receiving surface respectively. $R$ represents the distance between $ds_1$ and $ds_2$. Consider that the source surface is the freeform surface $\mathbf{f}(x', y') = (f_x, f_y, f_z)$ and the receiving surface is the target as shown in Fig. 2(b). Donations $\hat{\mathbf{N}}$, $\hat{\mathbf{N}}_t$ and $\hat{\mathbf{R}}$ represent the unit vector of freeform surface normal, target plane normal and the direction of rays respectively. After some mathematical deviations, Eq. (4) can be transformed to the following formula:

$$dE = \frac{L}{R^2}(z - f_z)[(\frac{\partial \mathbf{f}}{\partial x'} \times \frac{\partial \mathbf{f}}{\partial y'}) \cdot \hat{\mathbf{R}}] dx' dy'. \tag{3}$$

The irradiance of the target can be calculated by integrating Eq. (3). However, not every part on the freeform surface will contribute to the irradiance of a target point. Therefore, we implement a backward ray-tracing process to determine the radiance of rays emitted from the freeform surface to the target. As shown in Fig. 2(c), for each target point $\mathbf{t}(x, y)$, we reversely trace the rays emitted from $\mathbf{t}$ to the freeform surface $\mathbf{f}$. If the ray intersects within the range of the source, the radiance will be equal to that of this ray emerged from the light source in a lossless system. If the ray does not intersect with the source, the radiance of this ray will be defined as zero. Notice that we can directly trace a matrix of rays of one target point $(x, y)$ as shown in Fig. 2(d). Therefore, the radiance L in Eq. (5) can be regarded as a 4D function $L(x, y; x', y')$ which will be zero if the ray $(x, y; x', y')$ does not exist and be a certain value if the ray $(x, y; x', y')$ corresponds to a ray emerged from the light source.

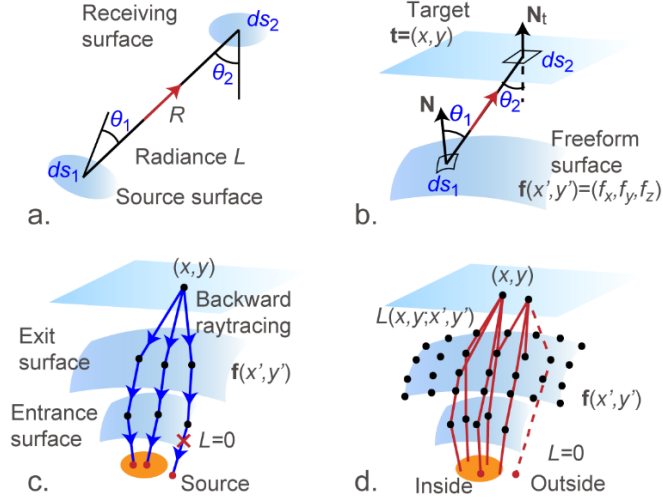

Fig. 2 Basic calculation of radiometry and freeform illumination system for extended sources

Based on the above considerations, the target irradiance can be calculated by an integral projection:

$$E(x, y) = \iint f(x, y; x', y') dx' dy', \quad (4)$$

where $f(x, y; x', y')$ is the integrand equals to the right-side of Eq. (3) removing the differential $dx'dy'$. Comparing the forms of Eq. (1) with Eq. (4), we conclude that the blur kernel function $h(x, y; x', y')$ can be formulated as:

$$h(x, y; x', y') = \frac{f(x, y; x', y')}{E_p(x', y')}, \quad (5)$$

where $E_p(x', y')$ is the irradiance distribution of an ideal source corresponding to the target point of the freeform lens.

**Spatially variant deconvolution**

The previous sections show that the illumination from extended sources can be formulated by a spatially variant convolution Eq. (1). Therefore, if the irradiance from a point source $E_p(x, y)$ has been given, we can calculate the freeform lens. And we can further obtain the blur kernel $h(x, y; x', y')$ using Eq. (5) and calculate the actual irradiance $E(x, y)$. In order to obtain a prescribed irradiance distribution, we replace the actual irradiance $E(x, y)$ by the desired irradiance $E_{desir}(x, y)$ and reversely solve a more reasonable target for ideal sources. When the new target irradiance for ideal sources is obtained, we can calculate a new freeform lens and estimate the blur kernel. We will iterate the above processes until obtaining a satisfactory result. In this section, we present the process of the calculation of the new target irradiance $E_p$ based on spatially variant deconvolution.

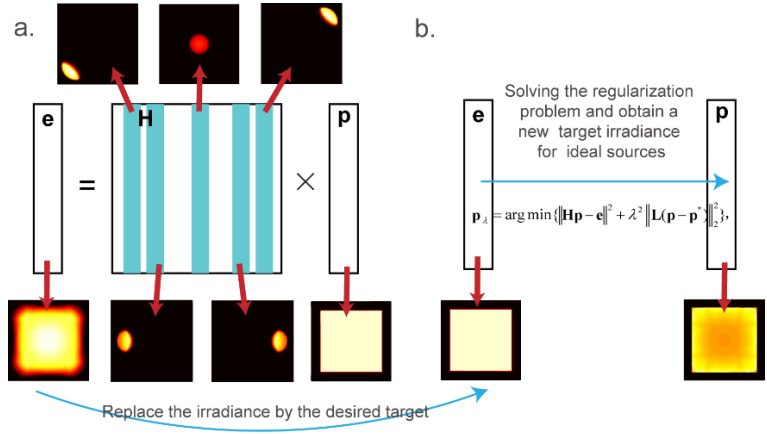

Fig. 3 The matrix form of spatially variant convolution and deconvolution in illumination system.

The spatially variant convolution Eq. (1) can be transformed into a matrix equation in the discrete form:

$$\mathbf{e} = \mathbf{Hp}, \tag{6}$$

where **e** denotes the irradiance of extended sources on point **t** and **p** represents the irradiance of ideal sources on point $(x, y)$. The matrix **H** is the discrete form of blur kernel $h(x, y; x', y')$. A more intuitive interpretation is presented in Fig. 3. Each column in matrix B denotes a blur pattern relating to a single point on target $(x', y')$ with irradiance $E_p(x', y')$. Integrating the blur pattern of all the points of target $(x', y')$, we will get the actual irradiance of extended sources.

The next step is to set the column vector **e** as the desired irradiance distribution and resolve a more reasonable irradiance distribution **p** of ideal sources. However, Eq. (6) is usually an ill-posed problem. This means that the direct solution of linear system $\mathbf{p} = \mathbf{H}^{-1}\mathbf{e}$ will lead to invalid result. Therefore, we transform Eq. (6) into a regularization problem [40]:

$$\mathbf{p}_\lambda = \arg\min\{\|\mathbf{Hp} - \mathbf{e}\|^2 + \lambda^2 \|\mathbf{L}(\mathbf{p} - \mathbf{p}^*)\|_2^2\}, \tag{7}$$

where the regularization parameter $l$ controls the weight given to the minimization of the side constraint relative to minimization of the residual norm. And the matrix **L** is typically either the identity matrix or a derivative operator. We use the function *csvd* and *tikhonov* in [38] to solve the iterated target irradiance of point sources. Once we have obtained the new target irradiance for ideal sources, we calculate a new freeform lens and implement a new loop presented in Fig. 1(c). A satisfactory result will be obtained usually in less than four iterations as shown in the next section.

### Results

To demonstrate the superiority of the proposed method, we designed two freeform lenses delivering different illumination patterns for far-field and near-field targets. All the designs use PMMA as the lens material and the light sources are Lambertian.

We first present a design example delivering a near-field square pattern to show the detailed processes of the proposed method. We define the Cartesian coordinates system with origin locating at the center of the light source and the z-axis being along with the normal of the plane source. The squared target distribution is located at a plane with $z_{target} = 100$ mm and side length equals 200 mm. We set a circular Lambertian source with its diameter equals to 19 mm. The Lambertian luminescent property means that the rays emerged from the light source with constant radiance $L$. The designed lenses should achieve a uniform target pattern and remain a highly compact configuration which means a low $h/d$ value.

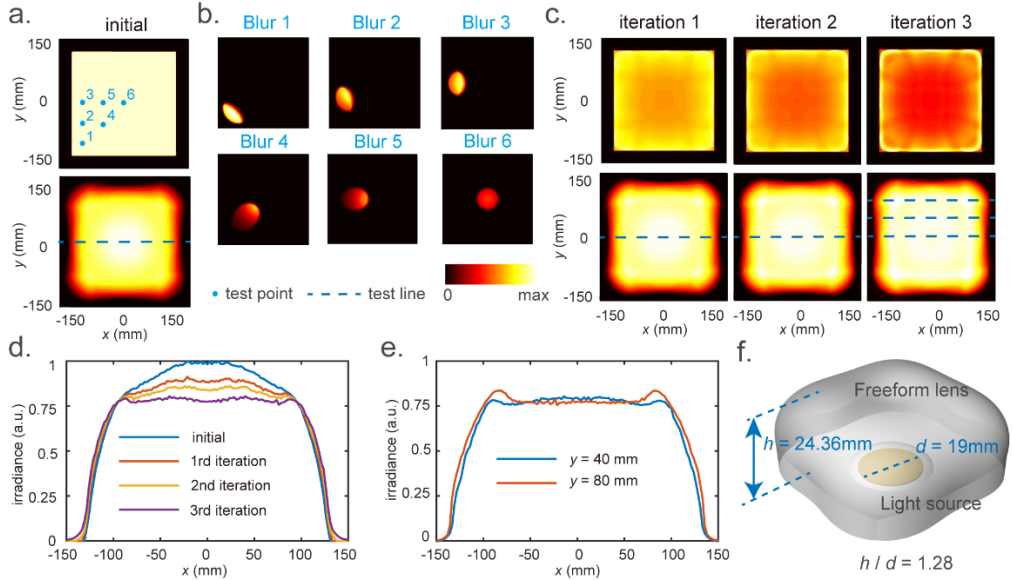

Fig. 4 The design of freeform illumination lens; (a) initial prescribed irradiance for the point source and the real irradiance distribution; (b) the blur patterns of six test points on ideal source target; (c) prescribed irradiances and real irradiances in each iteration; (d) profiles of irradiance on line $y = 0$ mm in each iteration; (e) profiles of irradiance on lines $y = 40$ mm and $y = 80$ mm; (e) the model of designed freeform lens with $h/d$=1.28.

The compact structure will lead to a serious blur effect on the target boundary. Therefore, we should set the size of target irradiance of ideal sources larger than the size of prescribed illumination pattern. As shown in the upward side of Fig. 4(a), the initial target for ideal sources is defined as a perfectly uniform squared pattern. We design the freeform lens based on the initial target with a nearly 180° collection angle of the virtual point source. The real irradiance distribution with the extended source is shown in the below picture of Fig. 4(a). The irradiance of ideal sources is blurred by the extension of the light source and the uniformity is degraded due to the blur effect. We then calculate the 4D blur kernel function $h(x, y; x', y')$ as a 2D matrix **H** expressed in Eq. (6). The Fig. 4(b) presents the blur patterns of each test point for an ideal source irradiance distribution. The new target irradiance of the ideal source can be calculated based on the spatially variant deconvolution process which is solving a regularization problem in our design framework shown in Section 2. The target irradiance distributions of the ideal source of each iteration are shown in Fig. 4(c). And the corresponding real irradiances with the extended source are presented in the below of Fig. 4(c). A satisfactory result is obtained in the 3$^{rd}$ iteration and the final irradiance distribution is shown in Fig. 4(c). The uniformity of the illumination area is 0.92 which is defined as the ratio of the average irradiance to the maximum irradiance $E_{aver}/E_{max}$. We plot the irradiance of two test lines $y = 40$ mm and $y = 80$ mm in Fig. 4(e). The lighting efficiency is defined as the flux captured by the target plane dividing by the total flux emitted from the light source. The lighting efficiency of the final freeform lens is 90.23% considering Fresnel losses, which is quite close to the value of theoretical limit 92.4% [22]. The designed freeform lens is presented in Fig. 4(e) with the maximum height equals to 24.36 mm, which shows an ultra-compact structure with $h/d = 1.28$. The design example has demonstrated the ability of this method to achieve ultra-efficient and compact freeform lenses with high optical performance for illumination applications.

The second design is to delivering a far-field rectangular pattern in off-axis configuration. The rectangular target distribution is located at a plane $z_{target}$ = 1000 mm with length 3000 mm and width 2000 mm. We set a square Lambertian source with its side length equals to 14 mm. We make the central point of the rectangular pattern moving a distance 300 mm. Fig. 5(a) shows the initial prescribed irradiance and the actual irradiance distribution. The blur patterns of six test points are presented in Fig 5(b). Only one iteration lead to a satisfactory result and the final prescribed irradiance as well as the actual irradiance are shown in Fig. 5(c). The profiles of $x$ = 0 mm and $y$ = 0 mm in each iteration are presented in Fig. 5(d). The uniformity of the rectangular area is 0.91 and lighting efficiency is 89.76% considering Fresnel loss. The final designed lens is shown in Fig. 5(e) with maximum height 41.43 mm, which presents a compact structure with $h/d$ = 2.09 in off-axis illumination design.

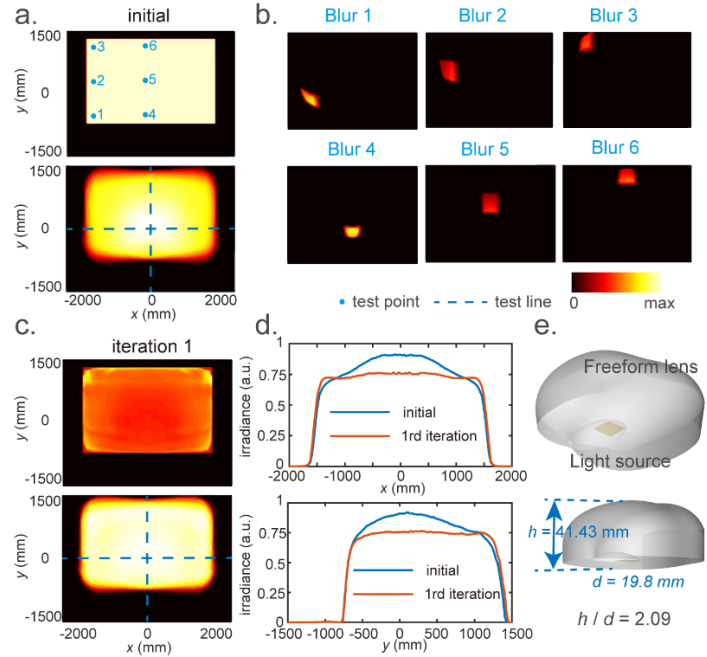

Fig. 5. The design of freeform illumination lens; (a) initial prescribed irradiance for the point source and the real irradiance distribution; (b) the blur patterns of six test points on ideal source target; (c) prescribed irradiances and real irradiances in each iteration; (d) profiles of irradiance on line $y$ = 0 mm in each iteration; (e) profiles of irradiance on lines $y$ = 40 mm and $y$ = 80 mm; (e) the model of designed freeform lens with $h/d$=1.28.

## Conclusion

We presented a spatially variant deconvolution method that can realize ultra-efficient, compact, and precise illumination design for extended sources. We connected the basic theory of radiometry and freeform optics with the mathematical model of spatially variant convolution, which produced a robust and generalized method for freeform illumination design. Two design examples were presented which have superior optical performances by comprehensively considering the efficiency, compactness, and accuracy of illumination control. The future work will focus on double surface design, complex pattern production, and multiple extended sources using the ideas of blur effects and spatially variant convolution, which we believe can forge a new path for illumination design and benefit a wide range of modern illumination systems.